\documentclass[10pt,journal,compsoc]{IEEEtran}

\ifCLASSOPTIONcompsoc
  \usepackage[nocompress]{cite}
\else
  \usepackage{cite}
\fi

\ifCLASSINFOpdf

\else

\fi

\usepackage{tikz} 
\usetikzlibrary{positioning}
\usetikzlibrary{arrows,decorations.pathmorphing,backgrounds,fit,petri}
\usepackage[autostyle]{csquotes}
\usepackage{placeins}
\usepackage{graphicx}
\usepackage{subcaption}
\usepackage[justification=centering]{caption}
\usepackage{amsthm}

\usepackage{amsmath}
\usepackage{enumitem}
\usepackage{multirow}
\usepackage{subcaption}
\usepackage{algorithm}% http://ctan.org/pkg/algorithm
\usepackage{algpseudocode}% http://ctan.org/pkg/algorithmicx
\usepackage{epstopdf}
\DeclareMathOperator*{\argmin}{argmin} % Jan Hlavacek
\algrenewcommand\algorithmicrequire{\textbf{Input:}}
\algrenewcommand\algorithmicensure{\textbf{Output:}}

\SetLabelAlign{LeftAlignWithIndent}{\hspace*{1.0ex}\makebox[1.25em][l]{#1}}

\begin{document}

\title{Reinforcement Learning for Optimal Load Distribution Sequencing in Resource-Sharing System}

\author{Fei~Wu,
Yang~Cao,
and~Thomas~Robertazzi,~\IEEEmembership{Fellow,~IEEE}% <-this % stops a space
\IEEEcompsocitemizethanks{\IEEEcompsocthanksitem Fei Wu, Yang Cao and T. Robertazzi are with the Department
of Electrical and Computer Engineering, Stony Brook Univesity, Stony Brook,
NY, 11794.\protect\\
E-mail: \{fei.wu,yang.cao,thomas.robertazzi\}@stonybrook.edu}}

\IEEEtitleabstractindextext{
\begin{abstract}
Divisible Load Theory (DLT) is a powerful tool for modeling divisible load problems in data-intensive systems. This paper studied an optimal divisible load distribution sequencing problem using a machine learning framework. The problem is to decide the optimal sequence to distribute divisible load to processors in order to achieve minimum finishing time. The scheduling is performed in a resource-sharing system where each physical processor is virtualized to multiple virtual processors. A reinforcement learning method called Multi-armed bandit (MAB) is used for our problem. We first provide a naive solution using the MAB algorithm and then several optimizations are performed. Various numerical tests are conducted. Our algorithm shows an increasing performance during the training progress and the global optimum will be acheived when the sample size is large enough.
\end{abstract}

}

\maketitle

\IEEEdisplaynontitleabstractindextext

\IEEEpeerreviewmaketitle

\IEEEraisesectionheading{\section{Introduction}\label{sec:introduction}}

\subsection{Background}
\IEEEPARstart{S}{cheduling} and resource management are important but challenge tasks for parallel processing systems. A major problem in scheduling is to decide the fraction of load for each processor to minimize the overall finishing time (i.e makespan). Such scheduling should be made under a consideration of network topology, load distribution manner, number of processors, processor's processing and link's communication speeds, etc. In this paper we will focus on the sequential distribution policy, where the load can be distributed to one processor at a time. Under such an assumption, the load distribution sequencing emerges as a crucial key to shorten the finishing time in heterogeneous system, where different processor has different computation/communication speed. \\
There are many ways to classify load sharing problems. One of them is to model the type of load as either indivisible or divisible. The indivisible load can not be partitioned and usually assigned to a single processor. However in modern data-intensive computing system, it is common to encounter large amount of similar data units, such as image processing, signal processing and so on [1,2]. Such load can be divided into parts of arbitrary sizes and processed in parallel. Scheduling for such data units can be solved efficiently by Divisible Load Theory (DLT).\\
The DLT was first studied by Cheng and Robertazzi in [3] and Agrawal and Jagadish in [23]. Since then, DLT has been utilized in various scheduling problems. For the sequential load distribution policy studied in this paper, DLT usually assumes that there is one or more control processors which hold the load at beginning. There are multiple worker processors which receive load from the control processor(s). The load distribution sequence is predefined and the control processor(s) transfers the load to worker processors one by one. The load amount for each processor is calculated from the optimality principle [4,5] that all the processors should finish processing at the same time. Our problem is to find the optimal predefined load distribution sequence to achieve the minimum finishing time.\\
\subsection{Optimal Sequencing in Resource-sharing System}
The optimal sequencing problem under the DLT framework has been studied in various papers. Optimal load distribution sequencing for heterogeneous system was studied in [6], where multi-round scheduling was taken into consideration. In [7], the optimal sequencing problem is studied in a single-level tree network with communication delays. Also, distribution sequencing of divisible load jobs with communication start-up costs is studied in [8]. In [8], a wide range of interconnection architectures of distributed computer systems is taken into consideration: a chain, a loop, a tree, a star of processors, a set of processors using shared buses and a hypercube of processors. Finally, optimal load distribution sequences for multi-level tree networks was studied in [9].\\
These previous works assume that the system is exclusively occupied by the divisible load job of interest, which indicates that the processor's full communication and computation power are entirely devoted to the divisible load job. However, such an assumption may not be true in the modern resource-sharing systems, where one processor can both communicate with multiple networks and process multiple jobs. Such resource-sharing technique and multi-tasking processors are commonly encountered in virtualized networks [10-12]. As a result, the extra communication and background jobs will take up system's communication and computation resources. The processor's processing and communication speeds for the divisible load job of our interest will be time-varying according the number of background jobs and extra communication links in the system.\\
\subsection{Our Contribution}
\begin{figure}[h!]
	\centering
	\includegraphics[width=0.5\textwidth]{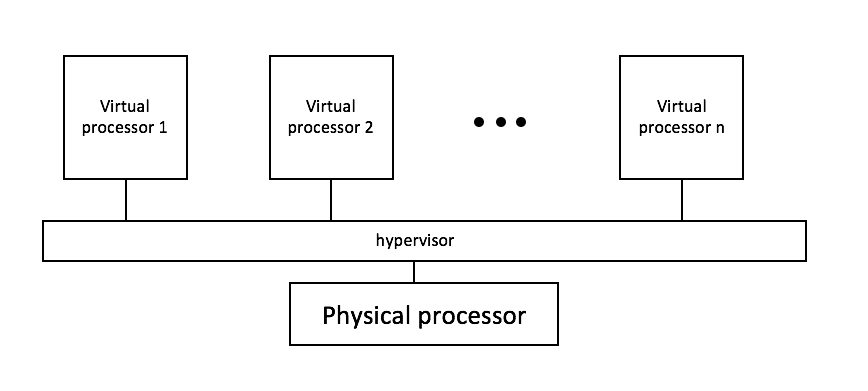}
	\caption{Processor virtualization}
	\label{fig:vir}
\end{figure}

This paper studied an optimal load distribution sequencing problem in resource-sharing systems. A single level tree network with heterogeneous channel is used as the network model in our system. A control processor holds the load at first. The load is then distributed to the worker processors in sequence. The communication and computation speeds of all processors are time-varying in our system due to virtualization. There is a hypervisor which is in charge of sharing the resources of the physical processor to the multiple virtual processors, as shown in Fig. \ref{fig:vir}.\\ 
Since the system is time-varying according to the randomly arrived background jobs, it is impossible to find an analytical solution for the optimal sequence. Instead, we perform a reinforcement learning method, multi-armed bandit (MAB), to train the system for the optimal sequencing. The MAB algorithm balances the explore and exploitation to minimize the regret. Such an algorithm is effective for rapid experimentation because it concentrates testing on actions that have the greatest potential power [13]. The empirical performance is good for MAB algorithms [14] and they are easy to apply. For each time a new divisible load job arrives at the control processor, MAB will help the user select a load distribution sequence. The system will learn the optimal sequence step by step. Thompson sampling (TS) is used in our MAB algorithm, which is a randomized Bayesian algorithm. The details will be discussed in section 3. 
\subsection{Organization}
The rest of this paper is organized as follows. We begin with time-varying scheduling preliminaries in Section 2. Next a reinforcement learning algorithm for optimal sequencing is introduced in Section 3. We first introduce the basic TS based MAB algorithm, then several optimization were studied for a better space and time complexity. The numerical test is in Section 4 and the conclusion is in Section 5.\\
The following notations are used in this paper:
\begin{enumerate} [leftmargin=1em,align=left]
	\item[$\kappa_{i}$] The partition of the entire divisible load that is assigned to processor $i$.
	\item[$\omega_{i}$] Inverse of processing speed of $ith$ processor when there is only one job.
	\item[$\omega_{i}(t)$] Inverse of time-varying processing speed of $ith$ processor applied to the divisible job at interest.
	\item[$\bar{\omega}_{i}$] Equivalent constant value of $\omega_{i}(t)$ during the processing time.
	\item[$T_{cp}$] Time to process the entire load when $\omega_{i} = 1$ for the $ith$ processor.
	\item[$z_i$] Inverse of channel speed when control processor is only communicating with $ith$ processor.
	\item[$z_{i}(t)$]  Inverse of time-varying channel speed applied to the divisible job at interest.
	\item[$\bar{z}_{i}$] Equivalent constant value of $z_{i}(t)$ when the control processor is communicating with the $ith$ worker processor
	\item[$Beta(\alpha,\beta)$] Beta distribution with parameter $\alpha,\beta$.
	\item[$T_{cm}$] Time to transmit the entire load when $z = 1$.
	\item[$T_{f}$] The finishing time of processing the entire load.
\end{enumerate}

\section{Preliminaries}
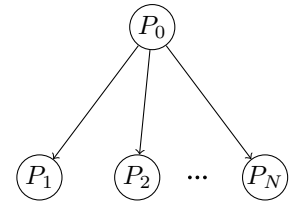
\begin{figure}[h!]
	\centering
	\begin{tikzpicture}
	[processor/.style = {circle,draw,inner sep=0pt,minimum size=6mm}]
	\node[processor] (p0) at (0,1.5) {$P_{0}$};
	\node[processor] (p1) at (-1.5,-0.5) {$P_{1}$};
	\node[processor] (p2) at (-0.2,-0.5) {$P_{2}$};
	\node[processor] (p3) at (1.5,-0.5) {$P_{N}$};
	\draw [->] (p0) to (p1);
	\draw [->] (p0) to (p2);
	\draw [->] (p0) to (p3);
	\filldraw [black] (0.5,-0.5) circle [radius=0.5pt];
	\filldraw [black] (0.6,-0.5) circle [radius=0.5pt];
	\filldraw [black] (0.7,-0.5) circle [radius=0.5pt];
	\end{tikzpicture}
	\caption{Single level tree network} 
	\label{fig:singleleveltree}
\end{figure}

The scheduling problem in resource-sharing system has been studied in [15] and [16]. In this section, we briefly introduce the problem formulation and solutions. This will be utilized in the later sections where we train the load distribution sequence.\\
Consider a single level tree network with N+1 nodes in Fig. \ref{fig:singleleveltree}.
The divisible load originally arrived at the control processor $P_{0}$. The control processor will divide the load into N+1 parts, which will be assigned to each processor in the system respectively. In this paper, it is assumed that the control processor can only communicate with one processor at a time. Due to the resource-sharing and virtualization, the processors' processing speed and communication speed are time-varying. According to the optimality principle [4,5], all processors should finish processing at the same moment. The system timing diagram is shown in Fig. \ref{fig:case2time}.\\
\begin{figure}[h!]
	\centering
	\begin{tikzpicture}
	%P0
	%time lane
	\draw[->] (-4.5,3) -- (3.5,3);
	%start lane
	\draw (-4.5,1.8) -- (-4.5,4.4);
	\draw (-4.8,3) node {$P_{0}$};
	%Communication blocks
	%\draw (-4.5,3) rectangle (-3.2,3.5);
	\draw (-4.5,3.5) -- (-4,3.5);
	\draw (-4,3.5) -- (-4,3.8);
	\draw (-4,3.8) -- (-2,3.8);
	\draw (-2,3.8) -- (-2,4.1);
	\draw (-2,4.1) -- (-0.5,4.1);
	\draw (-0.5,4.1) -- (-0.5,3.8);
	\draw (-0.5,3.8) -- (1.4,3.8);
	\draw (1.4,3.8) -- (1.4,3);
	\draw (-2.8,3.8) -- (-2.8,3);%T1
	\draw (-3.6,3.25) node {$\kappa_{1}\bar{z_{1}}T_{cm}$};
	\draw (-1.1,4.1) -- (-1.1,3);%T2
	\draw (-1.9,3.25) node {$\kappa_{2}\bar{z}_{2}T_{cm}$};
	\draw (-0.3,3.8) -- (-0.3,3);%TN
	\draw (0.6,3.25) node {$\kappa_{N}\bar{z}_{N}T_{cm}$};
	\draw[dashed] (-4.5,3.8) -- (-4,3.8);
	\draw[dashed] (-4.5,4.1) -- (-2,4.1);
	\draw (-4.8,3.5) node {$z_1$};
	\draw (-5.0,3.9) node {$z_1^h(2)$};
	\draw (-5.0,4.3) node {$z_2^h(3)$};
	\filldraw [black] (-0.8,3.25) circle [radius=0.5pt];
	\filldraw [black] (-0.7,3.25) circle [radius=0.5pt];
	\filldraw [black] (-0.6,3.25) circle [radius=0.5pt];
	\draw (-4.5,2.5) -- (-3.5,2.5);
	\draw (-3.5,2.5) -- (-3.5,2);
	\draw (-3.5,2) -- (-2.6,2);
	\draw (-2.6,2) -- (-2.6,2.5);
	\draw (-2.6,2.5) -- (-2,2.5);
	\draw (-2,2.5) -- (-2,2);
	\draw (-2,2) -- (1,2);
	\draw (1,2) -- (1,2.5);
	\draw (1,2.5) -- (2,2.5);
	\draw (2,2.5) -- (2,2);
	\draw (2,2) -- (3,2);
	\draw (3,2) -- (3,3);
	\draw[dashed] (-4.5,2) -- (-3.5,2);
	\draw (-4.8,2.5) node {$\omega_{0}$};
	\draw (-5.0,2) node {$\omega_{0}^h(2)$};
	\draw (-0.5,2.55) node {$\kappa_{0}\bar{\omega}_{0}T_{cp}$};
	\draw[->] (-4.5,1.2) -- (3.5,1.2);
	\draw (-4.5,0) -- (-4.5,1.5);
	\draw (-4.8,1.2) node {$P_{1}$};
	\draw (-2.8,1.2) -- (-2.8,0.7);
	\draw (-2.8,0.7) -- (-1.6,0.7);
	\draw (-1.6,0.7) -- (-1.6,0.2);
	\draw (-1.6,0.2) -- (1,0.2);
	\draw (1,0.2) -- (1,0.7);
	\draw (1,0.7) -- (3,0.7);
	\draw (3,0.7) -- (3,1.2);
	\draw[dashed] (-4.5,0.7) -- (-2.8,0.7);
	\draw[dashed] (-4.5,0.2) -- (-1.6,0.2);
	\draw (-4.8,0.7) node {$\omega_{1}$};
	\draw (-5.0,0.2) node {$\omega_{1}^h(2)$};
	\draw (-0.4,0.75) node {$\kappa_{1}\bar{\omega}_{1}T_{cp}$};
	\draw[dashed] (-2.8,3) -- (-2.8,1.2);
	\draw (-3,1.4) node {$T_{1}$};
	\draw[->] (-4.5,-0.6) -- (3.5,-0.6);
	\draw (-4.5,-1.8) -- (-4.5,-0.3);
	\draw (-4.8,-0.6) node {$P_{2}$};
	\draw (-1.1,-0.6) -- (-1.1,-1.6);
	\draw (-1.1,-1.6) -- (1.6,-1.6);
	\draw (1.6,-1.6) -- (1.6,-1.1);
	\draw (1.6,-1.1) -- (3,-1.1);
	\draw (3,-1.1) -- (3,-0.6);
	\draw[dashed] (-4.5,-1.6) -- (-1.1,-1.6);
	\draw[dashed] (-4.5,-1.1) -- (-1.1,-1.1);
	\draw (-4.8,-1.1) node {$\omega_{2}$};
	\draw (-5.0,-1.6) node {$\omega_{2}^h(2)$};
	\draw (0.5,-1.05) node {$\kappa_{2}\bar{\omega}_{2}T_{cp}$};
	\draw[dashed] (-1.1,3) -- (-1.1,1.2);
	\draw[dashed] (-1.1,0.2) -- (-1.1,-0.6);	
	\draw (-1.3,-0.4)node {$T_{2}$};
	\filldraw [black] (-3.2,-2.0) circle [radius=0.5pt];
	\filldraw [black] (-3.2,-2.2) circle [radius=0.5pt];
	\filldraw [black] (-3.2,-2.4) circle [radius=0.5pt];
	\draw[->] (-4.5,-2.9) -- (3.5,-2.9);
	\draw (-4.5,-3.9) -- (-4.5,-2.6);
	\draw (-4.8,-2.9) node {$P_{N}$};
	\draw (1.4,-2.9) rectangle (3,-3.4);
	\draw[dashed] (-4.5,-3.4) -- (1.4,-3.4);
	\draw (-4.8,-3.4) node {$\omega_{N}$};
	\draw (2.2,-3.15) node {$\kappa_{N}\bar{\omega}_{N}T_{cp}$};
	\draw[dashed] (1.4,2.5) -- (1.4,1.2);
	\draw[dashed] (1.4,0.7) -- (1.4,-0.6);
	\draw[dashed] (1.4,-1.6) -- (1.4,-2.9);
	
	\draw(1.2,-2.7)node {$T_{N}$};
	\draw[dashed] (3,2.5) -- (3,1.2);
	\draw[dashed] (3,0.7) -- (3,-0.6);
	\draw[dashed] (3,-1.1) -- (3,-2.9);
	\draw(3,3.2)node {$T_{f}$};
	\end{tikzpicture}
	\caption{Timing diagram for single level tree network with time-varying channel speed and computing speed } 
	\label{fig:case2time}
\end{figure}
In Fig. \ref{fig:case2time} $z_i^h(k)$ and $\omega_i^h(k)$ represents the inverse of communication and computation speed of $i$th processor when there is $k-1$ background jobs in it. The superscript $h$ means that it is controlled by the hypervisor. Steps are used to represent the arrival and departure of the background jobs, and the value of $\omega(t)$ and $z(t)$ are noted on the vertical axis. The communication is above the time axis and the computation is below the time axis. One can find that all processors finish processing at the same time $T_f$. The equivalent constant value of the time-varying inverse communication and computation speed from $T_m$ to $T_n$ for processor $i$ is calculated by:
\begin{subequations}
	\begin{align}
	\bar{\omega}_{i} = \frac{T_{n} - T_{m}}{\int_{T_{m}}^{T_{n}} \frac{1}{\omega_{i}(t)}dt}\\
	\bar{z}_{i} = \frac{T_{n} - T_{m}}{\int_{T_{m}}^{T_{n}} \frac{1}{z_i(t)}dt}
	\end{align}
\end{subequations}
Both a recursive algorithm and a simulation-based algorithm are introduced in [16] to solve the scheduling problem under different situations.
\section{Problem Formulation}
\subsection{Problem Setting}
The optimal sequencing problem can be defined as the selection of a sequence of load distribution in order to achieve the minimum finishing time (makespan). For each arrived divisible load job, the algorithm informs the user which sequence of load distribution should be selected to achieve lower finishing time. For time-invariant case, it has been proved [17] that the order of distributing load resulting in the shortest schedule is the order of nondecreasing inverse of communication speeds ($z_1 \leq z_2 \leq ... \leq z_N$). However, in the situation of resource-sharing and virtualization, one channel may be shared by multiple jobs, and the number of jobs is varying with time, which makes the channel speed time-varying. In such scenario, the optimal load distribution sequencing cannot be directly achieved.\\
A stochastic multi-armed bandit (MAB) algorithm is implemented to deal with the problem. Here we will briefly introduce the MAB framework [18]. We are given a slot machine with $N$ arms. For each time step t, one of the arms will be selected to be played and a reward will be observed. The reward is assumed to be a random value obtained from some fixed distribution. The random rewards are i.i.d for a specific arm along the time axis. The MAB algorithm will decide which arm to play at each time step t to achieve the maximum reward.\\
The optimal sequencing problem is now combined with a stochastic MAB framework. It is assumed that there are $N$ worker processors, thus there are totally $K = N!$ combinations of load distribution sequence, which is defined as $S_1, S_2,...,S_K$. In this case, it is noted as there are $N!$ arms. The algorithm proceeds in trials t = 1, 2, ..., T, where each trial represents an incoming divisible load job. On trial t, a sequence $S_t$ is selected and a schedule is made by the method described in the previous section. Such schedule is implemented to process the divisible load job arrived at trial t and finishing time $T_f(S_t,t)$ is observed. For a certain sequence $S_i$, its corresponding $T_f(S_i)$ can be modeled as a random variable as the reward in MAB framework. Let $S^{\star}(t)$ denote the optimal sequence at time t such that:
\begin{align}
S^{\star}(t) = \argmin_i E[T_f(S_i, t)]
\end{align}
So the objective of our algorithm is to minimize the cumulative regret for not selecting the optimal sequence over T trials:
\begin{align*}
& \text{minimize} 
& & \text{regret}\\
& \text{subject to} 
& & \text{regret} = \sum_{t=1}^{T} E[T_f(S(t), t)] - E[T_f(S^{\star}(t), t)]
\end{align*}
From now on,  arms and sequences will be used exchangeable.
\subsection{Thompson Sampling Based Multi-armed Bandit Algorithm}
Thompson sampling (TS) [19] is a common algorithm that addresses the exploration-exploitation dilemma in the multi-armed bandit problem. The selection is based on the Bayesian posterior distribution of the reward for each arm. The Bernoulli bandit problem is well studied [20], where the rewards are either 0 or 1. Furthermore, [18] introduced a method to map the reward in range $[0,1]$. \\
For Bernoulli bandits, TS often uses a Beta distribution to model the Bernoulli means. For a certain arm $i$, the reward is a Bernoulli random variable with mean $\mu_i$. The $\mu_i$ is assumed to be a random variable with a Beta($\alpha$, $\beta$) distribution. The Beta distribution turns out to be a convenient choice for updating the posterior distribution for Bernoulli trials. Here the details of updating the posterior distribution will be discussed.\\
For a certain arm i, the reward $R_i$ follows a Bernoulli distribution with mean $\mu_i$. The mean $\mu_i$ is also the probability that $R_i = 1$. The $\mu_i$ has a prior distribution $f(\mu_i) \sim Beta(\alpha_i, \beta_i)$. Now let's assume that one trial is performed at arm i and the reward is $R_t$. So the posterior distribution of $\mu_i$ given observed $R_t$ can be written as:
\begin{align*}
f(\mu_i|R_t) = \frac{f(R_t|\mu_i)f(\mu_i)}{\int f(R_t|\mu_i)f(\mu_i)d\mu_i}
\end{align*}
Given $R_t$ is a Bernoulli random variable, the probability density function of $R_t$ given $\mu_i$ can be written as:
\begin{align*}
f(R_t|\mu_i) = \mu_i^{R_t}(1-\mu_i)^{1-R_t}
\end{align*}
The posterior distribution of $\mu_i$ can be further deduced as:
\begin{align*}
f(\mu_i|R_t) &= \frac{\frac{\Gamma(\alpha_i + \beta_i)}{\Gamma(\alpha_i)\Gamma(\beta_i)}\mu_i^{\alpha_i-1 + R_t}(1-\mu_i)^{\beta_i - R_t}}{\int_{0}^{1}\frac{\Gamma(\alpha_i + \beta_i)}{\Gamma(\alpha_i)\Gamma(\beta_i)}\mu_i^{\alpha_i-1 + R_t}(1-\mu_i)^{\beta_i - R_t}d\mu_i}\\
&= \frac{\frac{\Gamma(\alpha_i + \beta_i)}{\Gamma(\alpha_i)\Gamma(\beta_i)}\mu_i^{\alpha_i-1 + R_t}(1-\mu_i)^{\beta_i - R_t}}{\frac{\Gamma(\alpha_i + \beta_i)}{\Gamma(\alpha_i)\Gamma(\beta_i)}\frac{\Gamma(\alpha_i+R_t)\Gamma(\beta_i - R_t + 1)}{\Gamma(\alpha_i+\beta_i+1)}}\\
&= \frac{\Gamma(\alpha_i+\beta_i+1)}{\Gamma(\alpha_i+R_t)\Gamma(\beta_i - R_t + 1)}\mu_i^{\alpha_i-1 + R_t}(1-\mu_i)^{\beta_i - R_t}
\end{align*}
which is also a Beta distribution with parameters $\tilde{\alpha_i} = \alpha_i+R_t \ and \ \tilde{\beta_i} = \beta_i - R_t + 1$. Since the reward $R_t$ can only be 0 or 1, so when updating the posterior distribution, it is simply done by adding $\alpha$ or $\beta$ with one depending on whether the Bernoulli trial deliver a 1 or 0. With higher $\alpha$, $\beta$, the Beta random variable is more concentrated around the mean.\\
In order to map the finishing time to a interval between 0 and 1, an upper bound of the finishing time $T_f^{max}(t)$ is obtained for each time step $t$. Such an upper bound can be the estimation of sequential processing time with a single processor. With $T_f^{max}(t)$, the parallel processing time can be mapped into an interval of $[0,1]$. In this way, each arm's finishing time $T_f(S_i)$ is modeled as a Beta distribution of parameter $\alpha_i$ and $\beta_i$. Then the Bernoulli TS algorithm can be modified to fit our situation. The details is described in Algorithm I.\\
\begin{algorithm}
	\begin{algorithmic}[1]
		\State For each arm $S_i$, i = 1, 2, ..., $N!$, set $\alpha_{S_i} = 0$, $\beta_{S_i} = 0$  		
		\For {t = 1,2,3,..}
		\State Find the $T_f^{max}(t)$ based on the estimate of sequential processing time with a single processor.
		\State For each arm $S_i$, i = 1, 2, ..., $N!$, sample $T_f(S_i, t)$ from Beta($\alpha_{S_i}$, $\beta_{S_i}$) distribution.
		\State Select the arm $S(t) = \argmin_i T_f(S_i, t)$.
		\State Perform the scheduling based on the selected arm (sequence) and observe the real finishing time $T_f(t)$.
		\State Set $\tilde{T_f(t)} = T_f(t) / T_f^{max}(t)$.
		\State Perform a Bernoulli trial with success probability $\tilde{T_f(t)}$ and observe output $r_t$.
		\State If $r_t = 1, \alpha_{S(t)} = \alpha_{S(t)} + 1$, else $\beta_{S(t))} = \beta_{S(t)} + 1$ 
		\EndFor	
	\end{algorithmic}
	\caption{Thompson sampling for optimal load distribution sequencing}
\end{algorithm}
The TS algorithm has a good property of balance in exploration-exploitation. For exploitation, every time step the algorithm tends to pick the arm with minimum expectation of finishing time. For exploration, instead of simply using the mean as the selection criterion, a random sample is drawn from the distribution, which makes it possible that the arms with higher mean are still able to be selected, in a low probability. The algorithm I is proved [18] to have a upper bound regret of $O((\sum_{i=2}^{N} \frac{1}{\mu_i - \mu_1})^2 lnT)$ in time $T$, where the first arm is assumed to be the optimal arm without loss of generality.  
\section{Algorithm Optimization}
The idea of Algorithm I is straightforward: sample the estimated finishing time from the prior distribution of each arm, pick the smallest one each time step. This method works well for the compact system where the number of processors is not very large. For the large systems with 10 processors or more, Algorithm I turns out to be inefficient. Due to the fact that Algorithm I needs to store the distribution parameter for each arm, the space complexity is $O(N!)$. Also, the arm selection step is $O(N!)$ time complexity since every arm could be visited. Such complexity makes scaling a very significant problem for Algorithm I when the system size is large.\\
To deal with the space and time complexity problem, two solutions are provided here. To reduce the space complexity, we redesign the representation of the load distribution. Thus the parameter space can be significantly condensed. For time complexity, two alternative searching algorithms are introduced, which are trade-offs between global optimality and time complexity. 
\subsection{Parameter Space Reformulation}  
In this section the parameter space is reformulated with a novel representation of the load sequence. Assume there are $N$ processors, indexed from $1$ to $N$, respectively. The sequence vector $S_V$ is defined as:
\begin{align}
S_V = [sv_1^T, sv_2^T, ... , sv_N^T]^T
\end{align}
where $sv_i, i = 1,2,...N$ are all $N \times 1$ vectors and thus $S_V$ is a $N^2 \times 1$ vector. Each subvector $sv_i$ represents the identity of $i$th receiving load processor. The subvector $sv_i$ has only one entry which equals to one, while the other entries are all equal to zero. For example, if the $j$th entry of $sv_i$ equals to one, this represents that the $i$th processor to receive the load from the control processor is the processor with the index $j$. Such design allows efficient and concise representation of the load distribution sequence. By this definition, the sequence vector of $S_i$ will be represented as $S_v(S_i)$.\\
Now, instead of assigning parameters for each sequence, weights are assigned for all entries of sequence vector $S_V$. The weight vector is thus defined as:
\begin{align}
W = [w_1, w_2, ... , w_{N^2}]^T
\end{align}
Here, vector $W$ shares exactly same length as vector $S_V$ and each entry of $W$ corresponds to each entry of $S_V$. Each entry $w_i$ in W vector is a random variable which follows Beta($\alpha_i$, $\beta_i$). For each time step $t$, $W(t)$ is sampled from its prior distribution. Instead of simply estimating the finishing time, we calculate the score for each sequence $S_v(S_i)$, which is defined as:
\begin{align}
Score(S_V(S_i), t) = S_V(S_i)^TW(t), \quad i = 1, 2, ..., N!
\end{align}
By adopting the score as the updating criteria in the algorithm, the Algorithm I can be modified as Algorithm II.\\
\begin{algorithm}
	\begin{algorithmic}[1]
		\State For each entry $w_i$ of $W$, set $\alpha_i = 0$, $\beta_i = 0$  		
		\For {t = 1,2,3,..}
		\State Find the $T_f^{max}(t)$ based on the estimate of sequential processing time with a single processor.
		\State Sample $W(t)$ from prior Beta($\alpha$, $\beta$) distribution.
		\State Select the arm $S(t) = \argmin_i S_V(S_i)^TW(t)$.
		\State Perform the scheduling based on the selected arm (sequence) and observe the real finishing time $T_f(t)$.
		\State Set $\tilde{T_f(t)} = T_f(t) / T_f^{max}(t)$.
		\State Perform a Bernoulli trial with success probability $\tilde{T_f(t)}$ and observe output $r_t$.
		\For {i = 1,2,3,...$N^2$}
		\State If $r_t = 1, \alpha_i = \alpha_i + S_V(S(t))[i] $, else $\beta_i = \beta_i + S_V(S(t))[i]$, where $S_V(S(t))[i]$ is the $i$th entry in $S_V(S(t))$
		\EndFor
		\EndFor	
		\State The optimal sequence $S^{\star} = \argmin_i S_V(S_i)^TW(t)$.
	\end{algorithmic}
	\caption{Thompson sampling for optimal load distribution sequencing}
\end{algorithm}
It is important that when updating W's the posterior distribution, only the entries which were used in the sequence vector $S_V(S(t))$ are updated. So we have to change the weight only if the corresponding position is valid, as can be found at step 10 in Algorithm II: the $\alpha$ and $\beta$ are only updated if it is 1 for the $i$th entry in $S_V(S(t))[i]$. The score is designed to be positive correlated to the finishing time. Such relationship can be found when updating the posterior distribution of the $W$ vector. For example when $r_t = 1$, algorithm increases $\alpha$ values of the entries of $W$ vector. Such positive incentive indicates an increase of $T_f(S(t))$ and may also results in an increase of the score for $S(t)$ in the next time step, which makes it less possible to be selected.\\
By such parameter space reformulation, the space complexity is now $O(N^2)$, which is a impressive improvement from $O(N!)$. However, the time complexity still remains $O(N!)$ since every sequence will be traversed to find the one with minimum score. 
\subsection{Time complexity optimization}
In this section two approximation algorithms are introduced to relax the time complexity problem. 
\subsubsection{Hill Climbing Algorithm for Searching}  
Instead of an exhaustive search of all arms to find the global minimum score, an alternative searching algorithm is performed to find the local optimum. Hill climbing optimization [21] is a mathematical optimization technique which attempts to find a better solution by changing a single element at a step of an random initialized solution, i.e. climbing the hill. Such optimization technique may result in a local optimum, but can be more efficient than the exhaustive searching.\\
To combine the hill climbing optimization with Algorithm II, we initialize the solution with an random arm $S_0$ and perform a hill climbing process. For each step, a subvector $sv_j$ is randomly picked to optimize, which corresponds to the $j$th position to receive load. By swapping the subvector $sv_j$ with all subvectors $sv_i, i = 1,2,...N$, the suboptimal arm can be found which will produce a minimum score. This is equivalent to swap the position of two processor in the system. We repeat this step $m$ times, where $m$ is a predefined iteration parameter. The searching can also be terminated before $m$ iterations if none of the subvectors is eligible to be optimized. In order to deal with the local optimum problem, the algorithm is performed from $K$ random start sequences instead of a single one. The details of hill climbing algorithm is described in Algorithm III.\\
\begin{algorithm}
	\begin{algorithmic}[1]
		\Require $W(t)$-weight for each position
		\For {i = 1,2,3,..,K}
		\State Pick a sequence $S_i(0)$ randomly as the initialization.
		\For {j = 1,2,3,..,m}
		\State Randomly pick a position $p_j$ to optimize.
		\State Swap all processor with processor at position $p_j$, find the one $\tilde{S}_i(j-1)$ that minimize $S_V(\tilde{S}_i(j-1))^TW(t)$
		\State $S_i(j) =\tilde{S}_i(j-1)$
		\EndFor
		\EndFor
		\State $S^\star = \argmin_i S_V(S_i(m))^TW(t)$\\
		\Return $S^\star$
	\end{algorithmic}
	\caption{Hill climbing algorithm for searching}
\end{algorithm}
A new TS algorithm can be achieved by replacing the step 5 in Algorithm II by Algorithm III. Since the swapping step in Algorithm III takes $O(N)$ time, the new TS algorithm only has a $O(KmN)$ time complexity, compared with the previous $O(N!)$ time complexity for exhaustive search.\\ 
\subsubsection{Batch Optimization}
The previous hill climbing algorithm reduced the time complexity from $O(N!)$ to $O(KmN)$. Such an improvement can be efficient sometimes, but for very large systems, the parameter $m$ and $N$ should be set large in order to avoid the local optimum problem, which will make the $O(KmN)$ time complexity also slow. Under such circumstance, pre-precessing to find the optimal sequence may cost too much time and thus impede the real divisible load processing.\\
For a very large system, adding one more processor at the end of all other processors won't affect the finishing time too much, even if that processor is a very fast processor [22]. This fact that the sequence of the processors at the front have higher influence on the finishing time than the ones at the end provides the intuition that if the front processors' sequences can be first processed. Under such intuition, a concept of batch optimization is introduced here. \\
To perform the batch optimization, a batch number is set as $b_n$, which means how much batch is utilized. Then the processors are randomly divided into $b_n$ groups, the processors within each group are randomly sequenced. Each of these groups is called a "batch". Now, instead of finding the sequence of all processors, we find the sequence of the batches. Because the processors' sequence inside each batch is predefined by random choice, the optimal sequence of batches can be trained by Algorithm II. After the optimal batch sequence is trained, we start to train the processors' sequence inside each batch. Since the sequence of the processors at the front have higher impact than the the ones at the end, the batches are trained in a sqeuence from the first to the last. Such a training policy allows maximum utilization of training samples.  If the number of processors inside of each batch is still too large, we divide them into batches and train the sequence of batches. Such an operation can be performed recursively until all the batches finish training. A detailed algorithm is described in Algorithm IV\\
\begin{algorithm}
	\begin{algorithmic}[1]
		\Require $b_n$-batch number, $b_s$-maximum number of processors that do not need to be divided into batches
		\If {processor number $\leq$ $b_s$}
		\State call Algorithm II for training the sequences of all processors
		\Else
		\State divide the processors into $b_n$ batches, call Algorithm II for training the sequences of all $b_n$ batches
		\For {each batch}
		\State repeat from step 1
		\EndFor
		\EndIf		
	\end{algorithmic}
	\caption{Batch optimization}
\end{algorithm}
The batch optimization's time complexity for one trial is $O(max(b_n!, b_s!))$, which can be very small if the $b_n, b_s$ are not set to be large. Compared with hill climbing algorithm, batch optimization is time-efficient for one trail, but may need more samples to train. Also, since the processors are randomly selected for each batch, it is possible that the batch optimization never reaches the global optimum. Note that if the number of processors is divisible by the batch number, the remainder will be evenly shared by the front batches. The number of processors inside each batch is not required to be exactly the same.
\section{Numerical Test}
In this section numerical tests are performed to evaluate our algorithms. In order to have a clear vision of the regret, the algorithms are first tested on the time-invariant cases, where the optimal arm is known beforehand. The exhaustive search and hill climbing are also compared in a compact system where the number of processors is less than 5. Finally, a general case is analyzed where the system is time-varying and relatively large.
\subsection{Time-invariant Case}
In this section, a time-invariant system is used to evaluate our algorithms. The finish time of each sequence can be explicitly calculated when the system is time-invariant. The time-varying case will be analyzed in the next section.
\subsubsection{General Performance Analysis}
\begin{figure}[h!]
	\centering
	\includegraphics[width=0.5\textwidth]{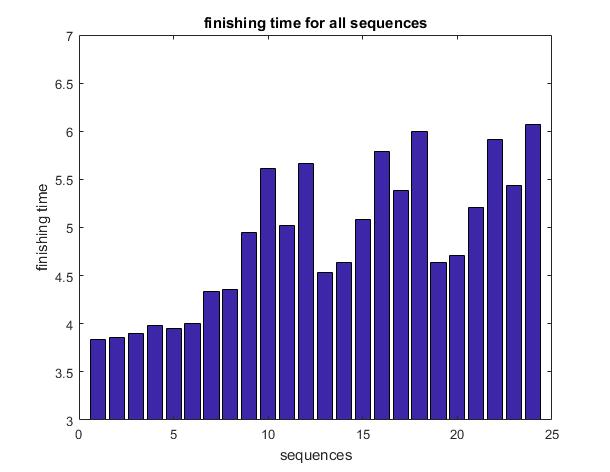}
	\caption{Finishing time for each sequences}
	\label{fig:1-1}
\end{figure}

\begin{figure}[h!]
	\centering
	\begin{subfigure}[b]{0.5\textwidth}
		\includegraphics[width=1\linewidth]{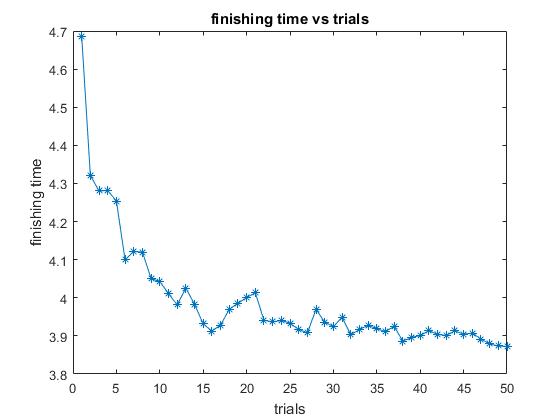}
		\caption{}
		\label{fig:1-2} 
	\end{subfigure}
	\begin{subfigure}[b]{0.5\textwidth}
		\includegraphics[width=1\linewidth]{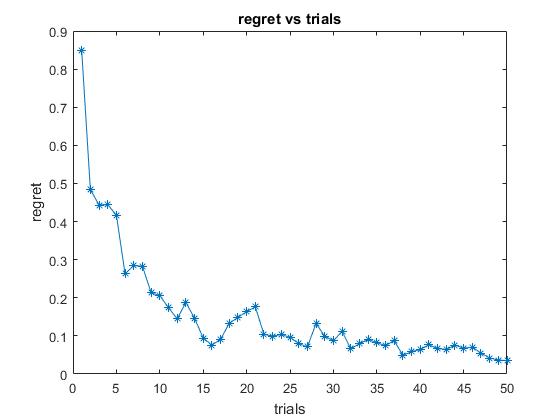}
		\caption{}
		\label{fig:1-3}
	\end{subfigure}
	\caption[]{For a time time-invariant system (a) Finishing time vs trials (b) Regret vs trials.}
\end{figure}

Assume there are one control processor and four worker processors. As a result, there are totally $4! = 24$ load distribution sequences. In this case it is assumed that the control processor does not process data. The inverse of the processing speeds for four worker processors are set to be $\omega_1 = 1, \omega_2 = 2, \omega_3 = 9, \omega_4 = 16$ and the inverse of communication speed $z_i$ is set equal to $\omega_i$. To ensure that the fact that the communication is usually faster than the computation, we set the $T_{cm} = 1$ and $T_{cp} = 4$. The finishing time for each sequence is shown in Fig. \ref{fig:1-1}.\\
There are totally 24 possible load distribution sequences. The fastest sequence yield a finishing time of 3.8370 time units, while the slowest one is 6.0725. Since the system is deterministic, i.e. there will not be any variation for the different trials. The cumulative regret during interval $T$ is simply calculated as:
\begin{align}
regret = \sum_{t=1}^{T} T_f(S(t),t) - T_f^\star  
\end{align}
where the $T_f(t)$ is the finishing time of the sequence $S(t)$ selected at time $t$ and $T_f^\star$ ids the finishing time of the optimal sequence. The $T_f^\star$ is a constant (in our case 3.8370) and $T_f(S(t),t)$ can never be smaller than $T_f^\star$. \\
The evaluation of Algorithm I is conducted by analysis the finishing time and regret as the algorithm is trained to select the optimal arm. In order to have a clear vision how the regret varies according to the trials, the average regret during interval $t_1$ and $t_2$ is defined as:
\begin{align}
regret = \frac{\sum_{t=t_1}^{t_2} T_f(S(t),t) - T_f^\star}{t_2-t_1+1}  
\end{align}
The average finishing time and regret is shown in Fig. \ref{fig:1-2} and Fig. \ref{fig:1-3}.\\
The Algorithm I is totally trained for 5000 trials. Both the $\alpha$ and $\beta$ are initialized as one for all arms. An average is taken for the finishing time and regret every 100 trials. i.e. every step on x axis is an average of 100 trials for Fig. \ref{fig:1-2} and Fig. \ref{fig:1-3}. It can be found that the finishing time and regret drops dramatically in the first 100 trials, where the algorithm is exploring all sequences to find the optimal sequence. After the first 100 trials the variation gradually decreases and the regret is approaching zero. In this phase the algorithm already find one or more sequences that are close to the optimal solution, but it is also possible to try other sequences with a small probability. One thing to notice is that even the system does not have enough samples to train the algorithm for the optimal solution, the algorithm can still yield decreasing finishing times over time.
\subsubsection{Exhaustive Search vs Hill Climbing} 
In order to resolve the scaling problem in time complexity and space complexity, a new parameter space is introduced in section 3. An approximating algorithm is also utilized to take place of exhaustive search: hill climbing. In this section the performance of Algorithm III is evaluated compared with the exhaustive search in Algorithm II.\\ 
The system parameters are the same as the last subsection. Since there are 4 worker processors, the weight vector $W$ is a $16 \times 1$ column vector and each element in $W$ vector is initialized as a random variable with $Beta(1,1)$. The $S_V$ vector is also a  $16 \times 1$ column vector. The hill climbing iteration $m$ is set to be 500. In order to resolve the local optimum problem, we use random start at Algorithm III. The number of random starts $K$ is set to be 20. \\
\begin{figure}[h!]
	\centering
	\begin{subfigure}[b]{0.5\textwidth}
		\includegraphics[width=1\linewidth]{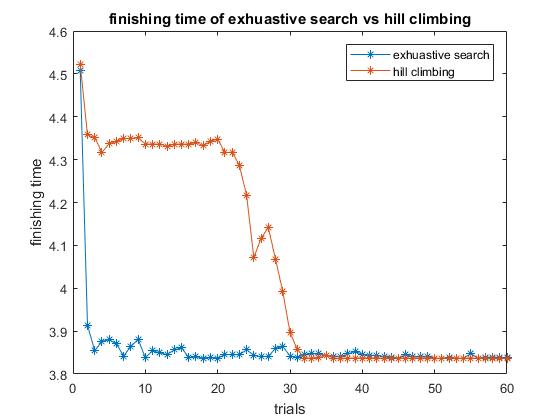}
		\caption{}
		\label{fig:2-1} 
	\end{subfigure}
	\begin{subfigure}[b]{0.5\textwidth}
		\includegraphics[width=1\linewidth]{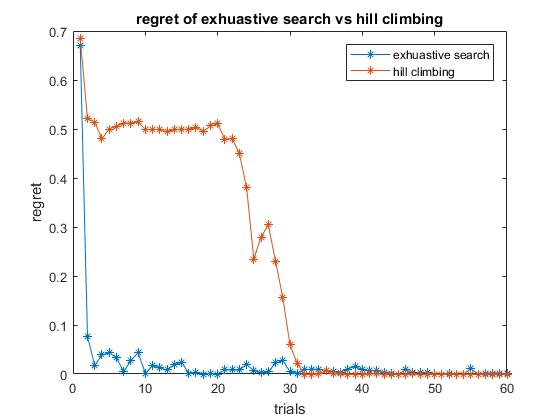}
		\caption{}
		\label{fig:2-2}
	\end{subfigure}
	\caption[]{Exhaustive search vs hill climbing (a) Finishing time (b) Regret.}
\end{figure}
Fig. \ref{fig:2-1} and Fig. \ref{fig:2-2} shows the performance of the exhaustive search and hill climbing algorithms. In this case the hill climbing converged to a local optimum while the exhaustive search converged to the global optimum when the sample size is small. When the sample size grows large, the hill climbing finally converged to the global optimum. This convergence correction is caused by the explore-exploitation balance property in Thompson sampling. Since the weight is also generated by drawing a sample from its distribution instead of just taking the mean, it is always possible to try solutions that seem not to be optimal due to randomness. Such randomness can help correct the error if the algorithm converges to a local optimal solution.\\
This case shows an experiment that the hill climbing converged to a local optimum. However, it is also possible that hill climbing can converge to the global optimum at first. The experiment result may vary due to randomness. Another thing to notice is that the new parameter space has a better performance than the old one when compare Fig. 6 with Fig. 5. This is because that the new parameter space give every position a weight to indicate which processor performs better in this position. Such design can be taken as having more features during the training process and thus results in a better solution.  
\subsection{General Time-varying System}
For the general time-varying case, the processor's processing (communication) speed will be time varying according to the number of background jobs (extra links). The physical processor will be virtualized to multiple virtual processors and a hypervisor will assign the communication/computation power of the physical processor to the virtual processors. For simplicity, it is assume that the hypervisor evenly distributes the communication/computation power to the virtual processors. In case there is no background jobs, the system parameters are set to be: $\omega_i  = 1 + i$ and $z_i = 1 + i, i = 1,2, .., N$ for totally N processors. Also, we set the $T_{cm} = 1$ and $T_{cp} = 4$ to make the communication is in general faster than computation.The number of background jobs (extra links) for each processor (channel) is simulated as a random number from 10 to 200 for totally 40 time units. As in the last section we take an average of 100 trials. The scheduling is performed by the algorithm in [16].\\
\subsubsection{Hill Climbing Optimization}
To perform the hill climbing optimization, the random initialization parameter $K$ and the search parameter are both set to be 100, which means for each trial, the best sequence is selected from the best of the hill climbing result of 100 random starts, each random start is optimized 100 times. The total comparison for hill climbing is $100 \times 100 = 10000$.\\
\begin{figure}[h!]
	\centering
	\includegraphics[width=0.5\textwidth]{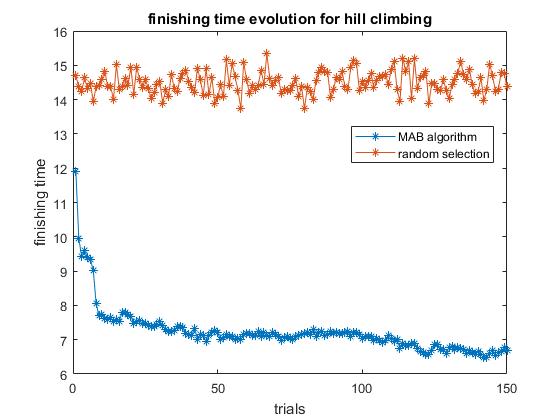}
	\caption{Finishing time for hill climbing optimization compared with random selection}
	\label{fig:3-1}
\end{figure}
Fig. \ref{fig:3-1} demonstrates the finishing time for a case of 20 processors and there are $20! = 2.4329 \times 10^{18}$ sequences in total. Under such scaling condition, Algorithm I is not a valid solution for this case.  Fig. \ref{fig:3-1} indicates that the algorithm can decrease the finishing time as the time grows. The local optimum problem is also appears in Fig. \ref{fig:3-1} as between the 15th to 25th averaged trials where there is small period of stable condition. Moreover, the finishing time at the end of Fig. \ref{fig:3-1} is not guaranteed to be the optimal solution, i.e. it is still possible to be locally optimal. Since the system is time-varying, the optimal solution may not be unique and may also be time-varying. It can be seen that the major drop of finishing time happens before the first 10 averaged trials, which indicates that even if the sample size is not large enough, our algorithm can still reduce the finishing time. 
\subsubsection{Batch Optimization}
Same as the hill climbing optimization, the number of processors is set to be 20. To perform the batch optimization, the batch number is set to be $b_n = 5$ and the maximum number of processors that do not need to be divided into batches $b_s$ is set to be 6. In this case, each batch has 4 processors so the inside batch optimization does not require further division. 
\begin{figure}[h!]
	\centering
	\begin{subfigure}[b]{0.5\textwidth}
		\includegraphics[width=1\linewidth]{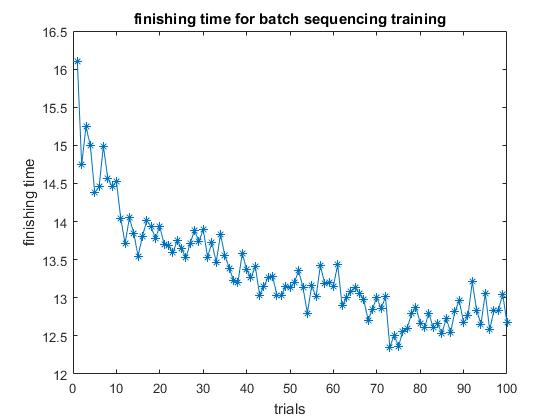}
		\caption{}
		\label{fig:3-2} 
	\end{subfigure}
	\begin{subfigure}[b]{0.5\textwidth}
		\includegraphics[width=1\linewidth]{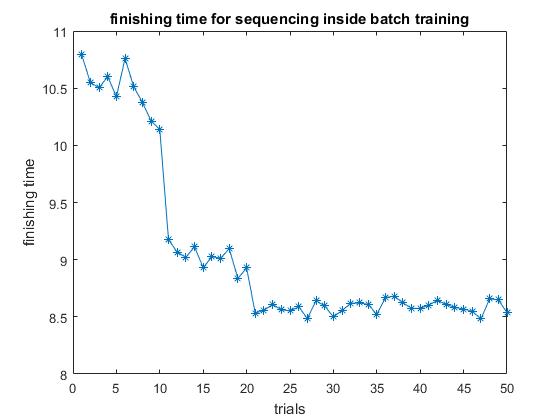}
		\caption{}
		\label{fig:3-3}
	\end{subfigure}
	\caption[]{batch optimization (a) batch sequencing (b) inside batch optimization.}
\end{figure}
Fig. \ref{fig:3-2} and Fig. \ref{fig:3-3} shows the 2 phases of batch optimization. The first phase is to find the optimal sequence of the batches and the second phase is to optimize the inside the batches. There are 5 batches and each batch has 4 processors. In Fig. \ref{fig:3-3} each batch takes 10 averaged trials to train the sequence. We can find that the finishing time dropping happens in the first 30 averaged trials, which means the first 3 batches. This meets our expectation since the processors at front have higher impact on the finishing time.\\
\begin{figure}[h!]
	\centering
	\includegraphics[width=0.5\textwidth]{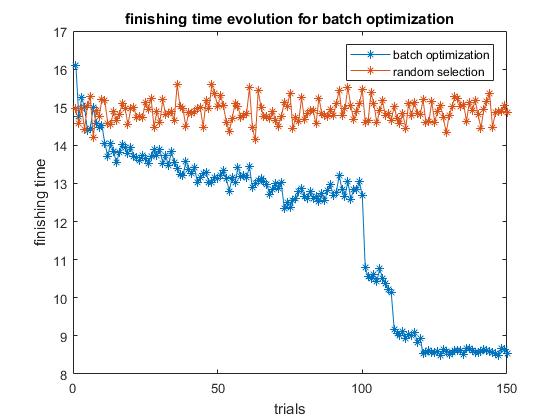}
	\caption{Finishing time for batch optimization compared with random selection}
	\label{fig:3-4}
\end{figure}
Fig. \ref{fig:3-4} shows the overall performance of batch optimization compared with random selection. Using the same parameters (sample size, number of processors) as the previous section of hill climbing optimization, it can be figure out that the training process for batch optimization is much slower than the hill climbing optimization. Such condition can be improved by dynamically adjust the sample size assigned for each training phase. For batch optimization, the total comparison for each trial is $max(b_n!,b_m!) = 120$, which is much smaller than the hill climbing optimization (which is 10000). 
\section{Conclusion}
This paper studied an optimal load distribution sequencing problem in a resource-sharing system. In the resource-sharing system the processors' communication and computation speeds are time-varying due to multi-tasking and virtualization. Thus the analytical solution can not be achieved. To deal with this issue, a reinforcement learning method was performed to train the optimal sequence using the multi-armed bandit algorithm. A Thompson sampling based multi-armed bandit algorithm was first introduced to train the optimal sequence. Then several optimization techniques were performed in order to decrease the time complexity and space complexity. The numerical test showed that our algorithm can deliver a continuing decreasing finishing time during the training progress and reach the global optimum if the sample size is large enough.\\
Future work for this research can be focused on other network topologies such as meshes or multi-level trees. It is also possible to combine the MAB algorithm with other optimization techniques such as genetic algorithm or random optimization instead of hill climbing. Dealing with the local optimum problem and achieving a better performance in large systems should also be further investigated.

\ifCLASSOPTIONcaptionsoff
  \newpage
\fi

\begin{IEEEbiography}[{\includegraphics[width=1in,height=1.25in,clip,keepaspectratio]{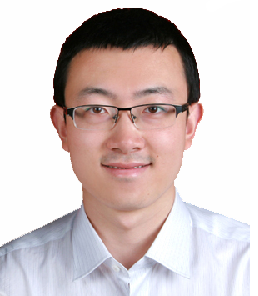}}]{Fei Wu}
	received the BE degree in information and telecommunication engineering from Xi'an Jiaotong University, Xi'an, China, in 2012, and the MS degree in electrical engineering from Stony Brook University, Stony Brook, New York, in 2013. He is currently working toward the PhD degree in electrical engineering at Stony Brook University. His research interests include scheduling, parallel processing, computer networks and virtualization. 
\end{IEEEbiography}
\begin{IEEEbiography}[{\includegraphics[width=1in,height=1.25in,clip,keepaspectratio]{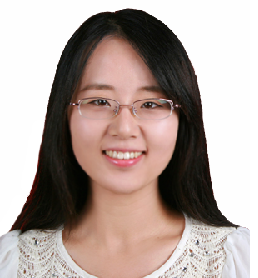}}]{Yang Cao}
	received the BE degree in Electrical Engineering and Automation from Northwestern Polytechnical University, Xi'an, China, in June 2012. She also received MS degree in Electrical Engineering from Stony Brook University, Stony Brook, New York, in December 2013. Currently she is working toward the PhD degree in Electrical Engineering at Stony Brook University. Her research interests include task scheduling and resource allocation in distributed systems, cloud networks,  data centers, etc.
\end{IEEEbiography}
\begin{IEEEbiography}[{\includegraphics[width=1in,height=1.25in,clip,keepaspectratio]{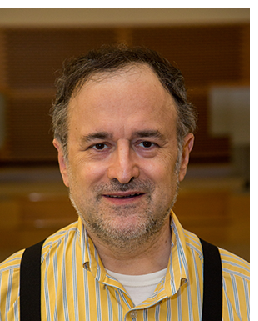}}]{Thomas G. Robertazzi}
	received the BEE degree from Cooper Union, New York, in 1977 and the PhD degree from Princeton University, Princeton, New Jersey, in 1981. He is presently a professor in the Department of Electrical and Computer Engineering, Stony Brook University, Stony Brook, New York. He has published extensively in the areas of parallel processing scheduling, telecommunications and performance evaluation. He has also authored, co-authored or edited six books in the areas of networking, performance evaluation, scheduling and network planning. He is a fellow of the IEEE and since 2008 co-chair of the Stony Brook University Senate Research Committee.
\end{IEEEbiography}

\end{document}